\documentclass[aps,12pt,prd,preprintnumbers,showpacs,nofootinbib]{revtex4}

\usepackage{amsmath}
\usepackage{amsfonts}

\newcommand{\pop}[1]{\frac{\partial}{\partial #1}}

\newcommand{\ca}[1]{{\cal #1}}
\newcommand{\SO}[1]{${\rm SO}(#1)$}
\newcommand{\vergule}{\,\,\,\,\,\,\,\,}

\newcommand{\be}{\begin{eqnarray}}
\newcommand{\ee}{\end{eqnarray}}
\newcommand{\nn}{\nonumber}

\newcommand{\ket}[1]{\left| #1 \right>}
\newcommand{\D}{{\cal D}}
\newcommand{\F}{{\cal F}}
\newcommand{\A}{{\cal A}}

\begin{document}

\preprint{ITEP-TH-52/09}

\title{SELF-DUALITY AND SUPERSYMMETRY}

\author{Maxim Konyushikhin}

\affiliation{SUBATECH, Universit\'e de Nantes, 4 rue Alfred Kastler,BP 20722, Nantes 44307, France}
\affiliation{Institute for Theoretical and Experimental Physics, B. Cheremushkinskaya 25, Moscow 117259, Russia}

\author{Andrei V. Smilga}

\affiliation{SUBATECH, Universit\'e de Nantes, 4 rue Alfred Kastler,BP 20722, Nantes 44307, France}
\affiliation{Institute for Theoretical and Experimental Physics, B. Cheremushkinskaya 25, Moscow 117259, Russia}

\begin{abstract}
We observe that the Hamiltonian $H =\, /\!\!\!\!{ \D}^2$, where $\,/\!\!\!\!{     \D}$ 
is the flat $4d$ Dirac operator in a self-dual gauge background, is supersymmetric, admitting 4 different real supercharges.
A generalization of this model to the motion on a curved conformally flat $4d$ manifold exists.
For an Abelian self-dual background, the corresponding Lagrangian
can be derived from known harmonic superspace expressions. 
\end{abstract}


\maketitle

 \section{Introduction}
The main purpose of this paper is to present a simple supersymmetric 
quantum mechanical model which, surprisingly, did not attract  
much attention so far. 

It was known since some time that one can treat the problem of the motion
of a fermion  
on an even-dimensional manifold with an arbitrary gauge field background 
as a supersymmetric one such that, e.g.,
the Atiyah-Singer index of a Dirac operator can be interpreted as the 
Witten index of a certain 
supersymmetric Hamiltonian \cite{Gaume}.
Our remark is that if the gauge field is self-dual and the $4d$ metric is
flat, the system enjoys an extended supersymmetry with two pairs 
of supercharges. A similar ${\cal N}=2$ supersymmetric
\footnote{${\cal N}$ counts the number of complex supercharges.}
 system can be written for conformally flat 4d manifolds, 
though supercharges in this case are not related to $\,/\!\!\!\!{     \D}$, and the Hamiltonian does not coincide with 
$\, /\!\!\!\!{ \D}^2$.

In Sect.~\ref{sect2}, we present the model. In Sect.~\ref{sect3}, we analyze in some more details its simplest version 
 (flat metric and constant self-dual Abelian field density) and derive the spectrum. In Sect.~\ref{sect4}, we derive the component
Lagrangian from a certain harmonic superspace (HSS) \cite{Galperin:2001uw} action suggested in  Ref.~\cite{Ivanov:2003nk}.

\section{Fermions in $4d$ self-dual background}\label{sect2}

Consider the Dirac operator in flat $4d$  Euclidean space
\begin{equation}
\label{gamE}
/\!\!\! \!\D \ =\ \sum_{\mu=0,1,2,3} \D_\mu \gamma_\mu\ ,
\end{equation}
where $\D_\mu = \partial_\mu - i {\cal A}_\mu$ and $\gamma_\mu$ are Euclidean anti-Hermitian gamma--matrices,
\begin{equation}\label{eq_gamma4}
\gamma_\mu = \left( \begin{array}{cc} 0  & -\sigma^\dagger_\mu \\
                                      \sigma_\mu & 0
 \end{array} \right),\vergule
\left\{\gamma_\mu,\gamma_\nu\right\}=-2\delta_{\mu\nu}\ ,
\end{equation}
with $(\sigma_\mu)_{\alpha \dot\beta} = \left\{i, \vec{\sigma} \right\}_{\alpha\dot\beta }$ and
$( \sigma^\dagger_\mu)^{\dot\beta \alpha} = \left\{-i, \vec{\sigma} \right\}^{\dot\beta\alpha}$ ($\vec\sigma$ are ordinary Pauli matrices). The indices are raised and lowered, as usual,
with antisymmetric Levi-Civita tensor $\varepsilon_{\alpha\beta} = \varepsilon_{\dot\alpha\dot\beta} = -\varepsilon^{\alpha\beta} = -\varepsilon^{\dot\alpha\dot\beta}$, $\varepsilon_{12}=1$.
(These are more or less the conventions of \cite{Wess} rotated to Euclidean space.) The Hamiltonians 
we are going to construct enjoy ${\rm SO}(4) = {\rm SU}(2) \times {\rm SU}(2)$ covariance such that the undotted spinor index refers
to the first ${\rm SU}(2)$ factor, while the dotted one to the second.   
 The matrices $\sigma_\mu,\ \sigma^\dagger_\mu$ satisfy the identities
\begin{equation}\label{eq_sigma}
\begin{array}{c}
  \sigma_\mu\sigma^\dagger_\nu+\sigma_\nu\sigma^\dagger_\mu= 
 \sigma^\dagger_\mu\sigma_\nu+\sigma^\dagger_\nu\sigma_\mu
=2\delta_{\mu\nu}, \\
  \sigma^\dagger_{\mu}\sigma_{\nu} - \sigma^\dagger_{\nu}\sigma_{\mu}
=2i\,\eta_{\mu\nu}^a \sigma_a, \\
  \sigma_{\mu}\sigma^\dagger_{\nu} - \sigma_{\nu}\sigma^\dagger_{\mu} =
2i\,\bar\eta_{\mu\nu}^a \sigma_a,
\end{array}
\end{equation}
where $\eta_{\mu\nu}^a$, $\bar\eta_{\mu\nu}^a$ are the  't~Hooft symbols,
\begin{equation}\label{tHooft}
\eta^a_{ij} = \bar\eta^a_{ij} = \varepsilon_{aij},\ \ \eta^a_{i0} = - \eta^a_{0i} = \bar\eta^a_{0i} =  -\bar\eta^a_{i0} =  \delta_{ai}
\end{equation}
($\sigma_a$ -- Pauli matrices, indices $a$, $i$, $j$ run from 1 to 3).
They are self-dual (anti-self-dual),
\begin{equation}
  \eta_{\mu\nu}^a=\frac{1}{2}\varepsilon_{\mu\nu\rho\lambda}\eta_{\rho\lambda}^a,\vergule
  \bar\eta_{\mu\nu}^a=-\frac{1}{2}\varepsilon_{\mu\nu\rho\lambda}\bar\eta_{\rho\lambda}^a,
\end{equation}
with the convention $\varepsilon_{0123} = -1$.
Another useful identity is
\begin{equation}\label{transpose}
\sigma_2 \sigma_\mu^T \sigma_2 = \ -\sigma^\dagger_\mu  \ .
\end{equation}

Consider the operator 
\begin{equation}
\label{HD2}
H =  \frac 12\, /\!\!\!\!\D^2  \ =\ - \frac 12 \D^2  - \frac i4  {\cal F}_{\mu\nu} \gamma_{\mu} \gamma_{\nu}
 \ ,
\end{equation}
where ${\cal F}_{\mu\nu}=\partial_\mu{\cal A}_\nu-\partial_\nu {\cal A}_\mu-i\left[{\cal A}_\mu, {\cal A}_\nu\right]$ is the field strength.
It is well known  that nonzero eigenvalues of the Euclidean Dirac operator come in pairs
$(-\lambda, \lambda)$ and hence the spectrum of the Hamiltonian $H$ is double-degenerate for all excited states. 
This means that, for any external field ${\cal A}_\mu$, this 
Hamiltonian is supersymmetric \cite{Gaume}  admitting two different anticommuting real supercharges: $/\!\!\!\!\D$ 
and $i/\!\!\!\! \D \gamma_5$ 
($\gamma_5 = \gamma_0 \gamma_1 \gamma_2 \gamma_3$).  Suppose now that the background field is self-dual,
\begin{equation}
\F_{\mu\nu} = \frac 12 \varepsilon_{\mu\nu\rho\delta} \F_{\rho\delta} \ \ \longleftrightarrow \ \ \F_{\mu\nu} = \eta^a_{\mu\nu} B_a\ .
\end{equation}
One can be easily convinced that in this case the Hamiltonian admits {\it four} different   Hermitian square roots $S_A$ that satisfy 
the extended supersymmetry algebra 
\begin{equation}
\label{N4}
\{S_A, S_B\} = 4\delta_{AB} H\ .
\end{equation}
One of the choices is 
\begin{equation}\label{matrS}
\begin{array}{l}
S_1  =   /\!\!\!\!\D =  \gamma_0 \D_0 + \gamma_1 \D_1 + \gamma_2 \D_2 + \gamma_3 \D_3 
 \ ,  \\
S_2   = \gamma_0 \D_3  + \gamma_1 \D_2 - \gamma_2 \D_1 - \gamma_3 \D_0   \ ,  \\
S_3 = \gamma_0 \D_2 - \gamma_1 \D_3 - \gamma_2 \D_0 + \gamma_3 \D_1   \ ,  \\
S_4 = \gamma_0 \D_1 -\gamma_1 \D_0 + \gamma_2 \D_3 - \gamma_3 \D_2  \ .
\end{array}
\end{equation}
Introducing the complex supercharges 
\begin{equation}
\label{QcherezS}
\begin{array}{c}
Q_1 = (S_1 - iS_2)/2,\vergule
Q_2 = (S_3 -  iS_4)/2, \\ 
\bar Q^1 = (S_1 + iS_2)/2,\vergule
\bar Q^2 = (S_3 + iS_4)/2,
\end{array}
\end{equation}
 we obtain the standard
 ${\cal N} = 2$ supersymmetry algebra
\footnote{
Note that, in contrast to $/\!\!\!\! \D$,  the operator $/\!\!\!\! \D \gamma_5$ is not expressed into a linear combination of $S_A$. 
In other words, the ${\cal N}=1$ supersymmetry algebra with the operators $/\!\!\!\! \D (1 \pm \gamma_5)$ is not a subalgebra of the 
${\cal N}=2$ algebra (\ref{eq_QQH}).
}
\begin{equation}\label{eq_QQH}
 \left\{Q_\alpha, Q_\beta\right\} = 0 ,\vergule
 \left\{Q_\alpha, \bar Q^\beta \right\}=2\delta_\alpha^\beta H.
\end{equation} 
Correspondingly, the excited spectrum of $H$ is four-fold degenerate, while the spectrum of $/\!\!\!\!\D$ 
consists of the quartets involving
two degenerate positive and two degenerate negative eigenvalues. 

The algebra (\ref{N4}) with supercharges (\ref{matrS}) holds for any self-dual field,  irrespectively of whether
it is Abelian or non-Abelian. Thus, 
the additional 2-fold degeneracy of the spectrum of the Dirac operator mentioned above should be there for a generic 
self-dual field. One particular example of a non-Abelian self-dual field is the instanton solution, where this degeneracy
was observed back in \cite{Jackiw} [see Eqs. (4.15) there].    

To make contact with the Lagrangian (and, especially, superfield) description, 
it is convenient to introduce holomorphic fermion variables,
which satisfy the standard anticommutation relations
\begin{equation}
\label{antikompsi}
\{ \psi_{\dot\alpha}, \psi_{\dot\beta} \} \ =\ \{ \bar\psi^{\dot\alpha}, \bar\psi^{\dot\beta} \} \ =\ 0, \vergule 
\{\bar \psi^{\dot\alpha}, \psi_{\dot\beta} \}
= \delta^{\dot\alpha}_{\dot\beta} \ .
\end{equation}
One of the possible choices is
\begin{eqnarray}
\label{psicherezgamma}
\psi_{\dot 1} = \frac {-\gamma_0 + i\gamma_3}2,\vergule 
\bar\psi^{\dot 1} = \frac {\gamma_0 + i\gamma_3}2,  \nn\ \\ 
\psi_{\dot 2} = \frac {\gamma_2 + i\gamma_1}2,\vergule 
\bar\psi^{\dot 2} = \frac {-\gamma_2 + i\gamma_1}2.\
\end{eqnarray}
Then two complex supercharges (\ref{QcherezS}) are expressed in a very simple way,
\begin{equation}
\label{eq_Q4}
\begin{array}{c}
  Q_\alpha= \left(\sigma_\mu  \bar\psi\right)_\alpha \left(\hat p_\mu-\ca A_\mu\right), \\ 
 \bar Q^\alpha= \left(\psi \sigma^\dagger_\mu  \right)^\alpha \left(\hat p_\mu-\ca A_\mu\right),
\end{array}
\end{equation}
with $\hat p_\mu = -i\partial_\mu$. 
 The Hamiltonian (\ref{HD2}) is expressed in these terms as 
\begin{equation}
\label{Hflatcherezpsi}
  H = \frac{1}{2} \left(\hat p_\mu-\ca A_\mu\right)^2 
     +  \frac{i}{4}\, {\cal F}_{\mu\nu}\,\psi \sigma^\dagger_{\mu} \sigma_{\nu}\bar\psi\ . 
\end{equation}
It is clear now why the spinor indices in Eq.(\ref{eq_QQH}) are undotted, while in Eq.(\ref{psicherezgamma}) they 
are dotted.
The supercharges are rotated by the first ${\rm SU}(2)$ and the variables $\psi_{\dot\alpha}$ by the second 
\footnote{Note that complex conjugation leaves the spinors in the same representation, the symmetry group here is ${\rm SO}(4)$ rather that ${\rm SO}(3,1)$.}. 
A careful distinction between two different ${\rm SU}(2)$ factors allows one to understand better the reason 
why the supercharges (\ref{eq_Q4}) satisfy the simple algebra (\ref{eq_QQH}) in a self-dual background.
The self-dual field density ${\cal F}$ carries in the spinor notation only dotted indices. Therefore any expression
involving ${\cal F}, \psi, \bar\psi$ is a scalar with respect to undotted ${\rm SU}(2)$. The only such scalar that can appear 
in the r.h.s. of the anticommutators of the supercharges $\{Q_\alpha, \bar Q^\beta\}$ is the structure which is 
proportional to $\delta_\alpha^\beta$, i.e. the Hamiltonian. No other operator is allowed.

In the {\it Abelian} case, the 
supercharges (\ref{eq_Q4}) and the Hamiltonian (\ref{Hflatcherezpsi}) are scalar operators not carrying matrix
indices anymore. This allows one to derive  the Lagrangian,
\begin{equation}
  \label{Lflat}
  L = 
    \frac{1}{2}\, \dot x_\mu\dot x_\mu
    +\ca A_\mu(x)\dot x_\mu
    +i{\bar\psi}^{\dot\alpha} \dot\psi_{\dot\alpha}
    -\frac{i}{4} {\cal F}_{\mu\nu}\,\psi\sigma^\dagger_{\mu}\sigma_{\nu}\bar\psi
  \ .
\end{equation}
 
In the non-Abelian case, the expressions (\ref{eq_Q4}, \ref{Hflatcherezpsi}) still keep their color matrix
structure, and one cannot derive the Lagrangian in a so straightforward way. One of the ways to handle the matrix structure is
 to introduce a set of color fermion variables (say, in the fundamental representation of the group) 
and impose
the extra constraint considering only the sector with unit fermion charge \cite{Gaume}. An alternative (non-Abelian) construction of the Lagrangian
is presented in \cite{IKS}, but in this paper we consider Lagrangians only for Abelian fields.

As will be demonstrated explicitly in Sect.~\ref{sect4}, the component Lagrangian (\ref{Lflat}) follows from the 
superfield action written earlier by Ivanov and Lechtenfeld  in the framework of harmonic superspace 
approach \cite{Ivanov:2003nk}.
We will see that one can naturally derive in this way a $\sigma$-model type generalization of the 
Lagrangian (\ref{Lflat}) describing
the motion over the manifold with nontrivial conformally flat metric $ds^2 = \left\{f(x)\right\}^{-2} dx_\mu dx_\mu$. 
It is written as follows
\begin{multline}\label{eq_action}
  L \ =\ 
    \frac{1}{2}f^{-2}\, \dot x_\mu\dot x_\mu
    +\ca A_\mu(x)\dot x^\mu
     +i{\bar\psi}^{\dot\alpha} \dot{\psi}_{\dot\alpha}
    - \frac{i}{4}f^2 {\cal F}_{\mu\nu}\,\psi\sigma^\dagger_{\mu}\sigma_{\nu}\bar\psi
\\
    +\frac{1}{4} \left\{3\left(\partial_\mu f\right)^2- f\partial^2 f \right\} \psi^4
     +\frac i2 f^{-1}\partial_\mu f \,\dot x_\nu\,
      \psi  \sigma^\dagger_{[\mu}\sigma_{\nu]} \bar\psi \ .
\end{multline} 

The corresponding (quantum) Noether supercharges and the Hamiltonian are
\begin{equation}\label{eq_Qf}
\begin{array}{l}
  Q_\alpha = f \left(\sigma_\mu \bar\psi\right)_\alpha \left(\hat p_\mu-\ca A_\mu\right)
-\psi_{\dot\gamma} \bar\psi^{\dot\gamma} \left(\sigma_\mu\bar\psi\right)_\alpha i\partial_\mu f,
\\[2mm]
  \bar Q^\alpha = \left(\psi\sigma^\dagger_\mu\right)^\alpha \left(\hat p_\mu-\ca A_\mu\right)f
+i\partial_\mu f \left(\psi\sigma^\dagger_\mu\right)^\alpha \cdot \psi_{\dot\gamma}\bar\psi^{\dot\gamma},
\end{array}
\end{equation}
\begin{multline}\label{eq_susyham}
  H=\frac{1}{2}f \left(\hat p_\mu-\ca A_\mu\right)^2 f
      +\frac{i}{4}f^2\, {\cal F}_{\mu\nu}\,\psi\sigma^\dagger_{\mu}\sigma_{\nu}\bar\psi
\\[2mm]
      - \frac 12 f i\partial_\mu f \,(\hat p_\nu-\ca A_\nu)\, \psi
	\sigma^\dagger_{[\mu} \sigma_{\nu]}\bar\psi
    + f\partial^2 f\left\{\psi_{\dot\gamma}\bar\psi^{\dot\gamma}-\frac{1}{2}\left(\psi_{\dot\gamma} \bar\psi^{\dot\gamma}\right)^2\right\}.
\end{multline}
On the other hand, one can explicitly calculate the anticommutators of the supercharges (\ref{eq_Qf}) 
 for any self-dual
\footnote{Anti-self-duality conditions are obtained when one interchanges  
$\sigma_\mu$ and $\sigma^\dagger_\mu$ in all the formulas. 
This is equivalent to the interchange of two spinor representations of \SO{4}.
}
field $\A_\mu(x)$, Abelian or non-Abelian, and verify that 
the algebra (\ref{eq_QQH}) holds.
While doing this, the use of the following Fierz identity
\begin{equation}
  \big(\bar\psi\sigma^\dagger_\mu\big)^\beta \big(\sigma_{\nu}\psi\big)_\alpha 
  - \big(\sigma_{\mu}\bar\psi\big)_\alpha \big(\psi\sigma^\dagger_{\nu}\big)^\beta = 
  \delta^\beta_\alpha \,\bar\psi \sigma^\dagger_\mu \sigma_\nu \psi \ ,
\end{equation}
which can be proven using (\ref{transpose}), is convenient.

Note that, with a nontrivial factor $f(x)$, the supercharges (\ref{eq_Qf}) have nothing to do with the Dirac
 operator $\,/\!\!\!\!{     \D}$ in a conformally flat background: the latter cannot be expressed as a 
linear combination of $Q_\alpha$ and $\bar Q^\alpha$. In addition, the Hamiltonian (\ref{eq_susyham}) does not coincide with $\,/\!\!\!\!{     \D}^2/2$.

The model (\ref{eq_action}-\ref{eq_susyham}) is a close relative to the model constructed in 
Ref.~\cite{Smilga:1986rb} (see Eqs. (30,31) there), which describes the motion on a {\it three}-dimensional conformally 
flat manifold in external magnetic field and a scalar potential. In fact, the latter model 
can be obtained from the former, if assuming that the metric
and the vector potential ${\cal A}_\mu \equiv (\Phi, \vec{\A})$ depend only on three spatial coordinates $x_i$. 
 If assuming further that the metric is flat, one is led to the Hamiltonian \cite{de Crombrugghe:1982un}
\begin{equation}\label{eq_3dham}
  H \ = \ \frac{1}{2}\left(\hat {\vec p}-\vec{\ca A}\right)^2+\frac{1}{2}\Phi^2+\vec\nabla \Phi \,\psi\vec\sigma\bar\psi,
\end{equation}
which is supersymmetric under the condition $ {\cal F}_{ij}  = \varepsilon_{ijk}
 \partial_k \Phi $ (the $3d$ reduction of the $4d$ self-duality condition). It was noticed in Ref.~\cite{Smilga:1986rb} 
that the effective Hamiltonian of a chiral supersymmetric electrodynamics
in finite spatial volume belongs to this class with $\Phi \propto 1/|\vec{A}|$. 
The vector potential $\vec{\cal A}(\vec{A})$ describes  
in this case a Dirac magnetic
monopole such that the Berry phase appears. The three dynamical variables $\vec{A}$ (do not confuse with curly $\vec{\cal A}$ !)
have in this case the meaning of 
the zero Fourier harmonic of the vector
potential in the original field theory. In the leading order, the metric is flat. 
When higher loop corrections are included, a (conformally flat !) metric on the moduli space $\{\vec{A}\}$ appears.

Performing the Hamiltonian reduction of Eq.~(\ref{eq_susyham}) with non-Abelian $\A_\mu$, a non-Abelian 
generalization of Eq.~(\ref{eq_3dham}) can easily be derived. It keeps the gauge structure of 
Eq.~(\ref{eq_3dham}) with matrix-valued $\vec \A$ and $\Phi$ satisfying the condition
 $\F_{ij}=\varepsilon_{ijk}\D_k \Phi$. Note that such Hamiltonian does {\sl not} coincide 
with the non-Abelian $3d$ Hamiltonian derived in Ref.~\cite{Bellucci:2009xr}.

\section{Constant  field}\label{sect3}

As an illustration, consider the system described by  the Hamiltonian (\ref{Hflatcherezpsi}) 
 in a constant self-dual Abelian background. The constant self-dual field strength $\F_{\mu\nu}
= \eta_{\mu\nu}^a B_a $ is parametrized by three independent components. Let us direct $B^a$ along the third axis, 
$B_a=(0,0,B)$, and  choose the gauge
\begin{equation}\label{eq_constfieldA}
	\A_0=Bx_3,\vergule
	\A_2=Bx_1,\vergule
	\A_1=\A_3=0.
\end{equation}
 The Hamiltonian (\ref{Hflatcherezpsi}) acquires the form
\begin{multline}\label{eq_constfield}
 H=\left\{\frac 12 \left(\hat p_0-Bx_3\right)^2+\frac 12 \hat p_3^2
				  +B \left(\chi_1\bar\chi^1-\frac 12\right)\right\}
\\
	+\left\{\frac 12 \left(\hat p_2-Bx_1\right)^2+\frac 12 \hat p_1^2
				  +B \left(\chi_2\bar\chi^2-\frac 12\right)\right\}.
\end{multline}
For convenience, we have introduced notations 
 $\chi_1=\bar\psi^{\dot 1}$, $\bar\chi^1=\psi_{\dot 1}$, $\chi_2=\psi_{\dot 2}$, $\bar\chi^2=\bar\psi^{\dot 2}$.
The  Hamiltonian is thus reduced to the sum $H_1 + H_2$ of two independent (acting in different Hilbert spaces)
supersymmetric Hamiltonians, each describing the 2-dimensional  motion of an electron in homogeneous orthogonal to the plane
magnetic field $\vec{B}$.   
The bosonic sector of each such Hamiltonian corresponds to the spin projection $\vec{s} \vec{B}/|\vec{B}| = -1/2$, and the fermionic
sector to the spin projection $\vec{s} \vec{B}/|\vec{B}| = 1/2$. 
This is the first and the simplest supersymmetric quantum problem ever
considered \cite{Landau}. 
 The energy levels for each Hamiltonian are 
$\varepsilon_i=B\left(n_i+\frac 12+s_i\right)$, $n_i\ge 0$ -- integers, $s_i=\pm \frac 12$. Each level of $H_i$ is doubly degenerate. Besides, there is an 
infinite degeneracy associated with the positions of the center of the orbit along the axes 1 and 3 that
are  proportional to the integrals of motion $p_2$ and $p_0$. The full spectrum  
\begin{equation}\label{eq_25}
 E=B\left(n_1+n_2+1+s_1+s_2\right)
\end{equation}
is thus 4-fold degenerate at each level (except for the state with $E=0$).

It might be instructive to explicitly associate this degeneracy with the action of supercharges (\ref{eq_Q4}).
Let us assume for definiteness $B > 0$. One can represent $Q_\alpha$ as
\begin{equation}
 Q_1=\sqrt{2B}\left(b\chi_1+a^\dagger\bar\chi^2\right),
\hskip 1cm
 Q_2=\sqrt{2B}\left(a\chi_1-b^\dagger\bar\chi^2\right)\ ,
\end{equation}
where $a^\dagger$, $b^\dagger$ and  $a$, $b$ are the creation and annihilation operators,
\begin{equation}
 a=\frac{1}{\sqrt{2B}}\left(\hat p_1 - iBx_1  \ + i p_2 \right),
\vergule
 b=\frac{1}{\sqrt{2B}}\left(\hat p_3 - iBx_3 \ + i p_0\right)\ ,
\end{equation}
\begin{equation}
 \left[a,a^\dagger\right]=1,\vergule
 \left[b,b^\dagger\right]=1. 
\end{equation}
In these notations, the Hamiltonian (\ref{eq_constfield}) takes a very simple form
\begin{equation}
 H=
	 B\left\{
		 a^\dagger a+b^\dagger b
		 +\chi_1\bar\chi^1+\chi_2\bar\chi^2
	 \right\}.
\end{equation}

Obviously, the energy levels of the Hamiltonian (\ref{eq_constfield}) are defined by two integrals of motion 
$p_{2,0}$, 
two oscillator excitation numbers $n_{1,2}$ and two spins $s_{1,2}$, as in Eq.~(\ref{eq_25}).
For each  $p_2, p_0$, there is a unique ground zero energy state $|0 \rangle$ 
annihilated by all supercharges. A quartet of excited states
can be  represented as 
 \begin{equation}
 \ket{n_1,n_2},\quad 
 Q_1^\dagger\ket{n_1,n_2},\quad
 Q_2^\dagger\ket{n_1,n_2},\quad
 Q_1^\dagger Q_2^\dagger\ket{n_1,n_2} \ ,
\end{equation}
where the state 
$$ \ket{n_1,n_2}\equiv\chi_1\cdot\left(a^\dagger\right)^{n_1}\left(b^\dagger\right)^{n_2}\ket{0}$$
of energy  $E=B(n_1+n_2+1)$ is annihilated by both $Q_1$ and $Q_2$. 

For each $p_2, p_0$, there are $N$ such quartets at the energy level $E = BN$. 




\section{From harmonic superspace to components}\label{sect4}

In this section, we derive  the Hamiltonian (\ref{eq_susyham}) in the HSS approach.
To make the paper self-consistent, we present in the Appendix its salient features and definitions in application to quantum 
mechanical problems.
 The relevant superfield action was written in \cite{Ivanov:2003nk}, and we show here that the corresponding component Lagrangian
coincides with (\ref{eq_action}). The corresponding supercharges (\ref{eq_Qf}) and the Hamiltonian (\ref{eq_susyham}) involve
an Abelian self-dual gauge field ${\cal A}_\mu (x)$. The non-Abelian case is treated in a separate publication \cite{IKS}.

Let us introduce a doublet of superfields $q^{+{\dot\alpha}}$ with charge +1 ($D^0 q^+ = q^+$) 
satisfying the constraints (\ref{eq_anal}).
The index ${\dot\alpha}$ is the fundamental representation index of an additional external group ${\rm SU}(2)$. 
The solution for these constraints in the analytical basis is [see Eq.(\ref{eq_q})] 
\begin{equation}\label{eq_qdot}
  q^{+\dot\alpha} = x^{\alpha\dot\alpha}(t_{\rm A})u^+_\alpha
  -2\theta^+\chi^{\dot\alpha}(t_{\rm A})-2\bar\theta^+\bar\chi'^{\dot\alpha}(t_{\rm A})
  -2i\theta^+\bar\theta^+\partial_{\rm A}x^{\alpha\dot\alpha}(t_{\rm A}) u^-_\alpha \ .
\end{equation}
We impose now the additional pseudoreality condition 
\begin{equation}
\label{reality}
  q^{+{\dot\alpha}}=\varepsilon^{{\dot\alpha}{\dot\beta}} \widetilde{q^{+{\dot\beta}}}\ ,
\end{equation}
the field $\widetilde{q}^+$ being defined in Eq.(\ref{eq_tildeq}).
It implies
\begin{equation}
  x^{\alpha{\dot\alpha}}=-\left(x_{\alpha{\dot\alpha}}\right)^* ,\vergule
  \bar\chi'^{\dot\alpha}=\left(\chi_{\dot\alpha}\right)^*\equiv\bar\chi^{\dot\alpha}\ .
\end{equation}

Let us go back  now to the central basis $\left\{t,\theta_\alpha,\bar\theta^\beta,u^\pm_\gamma\right\}$.
The solution can be presented as $q^{+{\dot\alpha}}=u^+_\alpha q^{\alpha{\dot\alpha}}$ where 
$q^{\alpha{\dot\alpha}}$ does not depend on $u^\pm_\alpha$ (the latter follows from the constraint $D^{++} q^{+{\dot\alpha}} = 0$
and the definition $D^{++} = u^+_\alpha \pop{u^-_\alpha}$).
It is convenient to go over to the $4d$ vector  notation, introducing 
\begin{equation}\label{eq_qqp}
  q_\mu= -\frac 12\left(\sigma_\mu\right)_{\alpha{\dot\alpha}} \, q^{\alpha{\dot\alpha}},\vergule
  q^{+{\dot\alpha}}=-q_\mu\left(\sigma^\dagger_\mu\right)^{{\dot\alpha} \alpha}u_\alpha^+ \ .
\end{equation}
 Now, $q_\mu$ is a vector with respect to
the group ${\rm SO}(4)={\rm SU}(2)\times{\rm SU}(2)$, with the first factor representing the 
 $\ca N=2$ R-symmetry group and the second one being the extra global ${\rm SU}(2)$ group rotating the dotted ``flavor'' indices.

Pseudoreality condition (\ref{reality}) implies that the superfield $q_\mu$ is { real}. The latter
 is expressed in components as follows, 
  \begin{equation}\label{eq_qmu}
  q_\mu=x_\mu
+\theta\sigma_\mu\chi+\bar\theta\sigma_\mu\bar\chi
-\frac{i}{2}\dot x_\nu\, \bar\theta \sigma_{[\mu}\sigma^\dagger_{\nu]}\theta
    +\frac{i}{2}\bar\theta\sigma_\mu\dot \chi \,\theta^2
    -\frac{i}{2}\theta\sigma_\mu\dot{\bar\chi} \,\bar\theta^2
  -\frac{1}{4}\ddot x_\mu \,\theta^4 \ ,
  \end{equation}
where $\theta^2 \equiv \theta^\alpha\theta_\alpha$, $\bar\theta^2 \equiv \bar\theta^\alpha\bar\theta_\alpha$, 
$\theta^4 \equiv \theta^2\bar\theta^2$.

The classical $\ca N=2$ SUSY invariant action for the superfield $q_\mu$ can now be written. It consists of two parts, $S=S_{\rm kin}+S_{\rm int}$. 
The kinetic part,
\begin{equation}
\label{Lkin}
  S_{\rm kin}=\int  dt\, d^4\theta du\, R'_{\rm kin}(q^{+{\dot\alpha}}, q^{-{\dot\beta}}, u^\pm_\gamma)=\int dt\,d^4\theta\, R_{\rm kin}(q_\mu),
\end{equation}
depends on  an arbitrary function $R_{\rm kin}(q_\mu)$. Plugging  (\ref{eq_qmu}) into (\ref{Lkin}) and adding/subtracting proper
  total derivatives, we obtain
\begin{equation}\label{eq_41}
  S_{\rm kin}=\int dt\left\{
    \frac{1}{2}g(x)\, \dot x_\mu\dot x_\mu
    +\frac{i}{2}g(x)\left(\bar\chi^{\dot\alpha}\dot\chi_{\dot\alpha}-\dot{\bar\chi}^{\dot\alpha}\chi_{\dot\alpha}\right)
  \right.
  \left.
    +\frac{1}{8}\partial^2 g(x) \,\chi^4
    -\frac{i}{4}\partial_\mu g(x) \,\dot x_\nu\,
      \chi\sigma^\dagger_{[\mu}\sigma_{\nu]}\bar\chi
  \right\},
\end{equation}
where $g(x)=\frac{1}{2}\partial^2_x R_{\rm kin}(x)$ and
$\chi^4=\chi^{\dot\alpha}\chi_{\dot\alpha}\bar\chi^{\dot\beta}\bar\chi_{\dot\beta}$.

To couple $x_\mu$ to an external gauge field, one should add  the interaction term $S_{\rm int}$ that represents an integral over 
{\it analytic} superspace,
\begin{equation}\label{eq_38}
  S_{\rm int}=\int dt\, du\, d\bar\theta^+ d\theta^+ R_{\rm int}^{++}\left(q^{+{\dot\alpha}}(t_{\rm A},\theta^+,\bar\theta^+),u^\pm_\gamma\right)
\ .
\end{equation}
 We choose $R_{\rm int}^{++}$ (it carries the charge 2) 
satisfying the condition $\widetilde R_{\rm int}^{++} = - R_{\rm int}^{++}$ [the involution operation $\widetilde X$ 
being defined in Eqs. (\ref{harminv}), (\ref{eq_involution})] such that the action (\ref{eq_38}) is real.

To do the integral over $\theta^+$ and $\bar\theta^+$,  introduce 
$x^{+{\dot\alpha}}=-x_\mu \left(\sigma^\dagger_\mu\right)^{{\dot\alpha} \alpha}u^+_\alpha\equiv x^{\alpha\dot\alpha}u^+_\alpha$
[see Eq.(\ref{eq_qqp})]. Then
\begin{equation}
\label{Lint}
  S_{\rm int}=
    \int dt\,du\left\{2i\left(\sigma^\dagger_\mu\right)^{{\dot\alpha} \alpha} \partial_{+\dot\alpha} R_{\rm int}^{++}\, u_\alpha^- \cdot \dot x_\mu
    -4\chi^{\dot\alpha}\bar\chi^{\dot\beta}\,\partial_{+\dot\alpha}\partial_{+\dot\beta} R_{\rm int}^{++}
    \right\}
\end{equation}
 with
\begin{equation}
  \partial_{+\dot\alpha} R_{\rm int}^{++} (x,u)\equiv
  \frac{\partial R_{\rm int}^{++} (x^{+{\dot\gamma}}, u^\pm_\gamma)}{\partial x^{+{\dot\alpha}}}.
\end{equation}
Now, define the gauge field,
\begin{equation}
\label{defA}
  \ca A_\mu(x) \equiv \int du\left\{2i\left(\sigma^\dagger_\mu\right)^{{\dot\alpha} \alpha} \partial_{+\dot\alpha} R_{\rm int}^{++}\, u_\alpha^-\right\}.
\end{equation}
As the action (\ref{Lint}) is real, the field $\ca A_\mu(x)$ is also real. It has zero divergence, $\partial_\mu \A_\mu=0$.

The field strength is expressed as 
\begin{equation} 
\label{defF}
  \F_{\mu\nu} = \partial_\mu \ca A_\nu-\partial_\nu \ca A_\mu
  = -2 \eta_{\mu\nu}^a  \int du\, \partial_{+\dot\alpha}\partial_{+\dot\beta} R_{\rm int}^{++}\, 
\varepsilon^{{\dot\alpha}{\dot\gamma}}
    \left( \sigma_a  \right)^{\!{\dot\beta}}_{\,\,{\dot\gamma}}
\end{equation}
(the identities (\ref{eq_sigma}) were used). It is obviously self-dual. 
 With the definitions (\ref{defA}) and (\ref{defF}) in hand, one can represent the interaction term (\ref{Lint}) as
\begin{equation}
\label{43}
  S_{\rm int}=\int dt\left\{\ca A_\mu(x)\dot x_\mu
  -\frac{i}{4}\F_{\mu\nu}\,\chi\sigma^\dagger_{\mu}\sigma_{\nu}\bar\chi\right\}.
\end{equation}
Adding this to the kinetic term in (\ref{eq_41}) [where one can get rid of the 
factor $g(x)$ in the fermion kinetic term by introducing canonically conjugated
  $\psi_{\dot\alpha}=f^{-1}(x)\chi_{\dot\alpha}, \ \bar\psi^{\dot\alpha} = f^{-1}(x) \bar\chi^{\dot\alpha} $
with $f(x) = g^{-1/2}(x)$],
one can explicitly check that the Lagrangian $L = L_{\rm kin}+L_{\rm int}$ coincides, up to a total derivative, with (\ref{eq_action}).
The action is invariant under supersymmetry transformations,
\begin{equation}
\begin{array}{c}
  x_\mu\rightarrow x_\mu
  +f\epsilon\sigma_\mu\psi
  +f\bar\epsilon\sigma_\mu\bar\psi,
\\[3mm]
  f\psi_{\dot\alpha}\rightarrow f\psi_{\dot\alpha}
  +i\dot x_\mu \left(\bar\epsilon\sigma_\mu\right)_{{\dot\alpha}},
\\[3mm]
  f\bar\psi^{\dot\alpha}\rightarrow f\bar\psi^{\dot\alpha}
  -i\dot x_\mu \left(\sigma^\dagger_\mu\epsilon\right)^{{\dot\alpha}}.
\end{array}
\end{equation}
 The Noether classical supercharges expressed in terms of  $\psi_{\dot\alpha}, \bar\psi^{\dot\alpha}$, $x_\mu$ and their canonical momenta,
\begin{equation}
\label{pmu}
p_\mu \ =\ f^{-2}\dot x_\mu + {\cal A}_\mu - \frac i2 f^{-1} \partial_\nu f\,\psi \sigma^\dagger_{[\mu} \sigma_{\nu]} \bar \psi\ ,
\end{equation}
are 
\begin{equation}
\begin{array}{ccl}
 Q_\alpha &=& f \left(\sigma_\mu \bar\psi\right)_\alpha \left( p_\mu-\ca A_\mu\right)
-i  \partial_\mu f \psi_{\dot\gamma} \bar\psi^{\dot\gamma} \left(\sigma_\mu\bar\psi\right)_\alpha \nonumber, \\
 \bar Q^\alpha &=& \mbox{[complex conjugate]} \ .
\end{array}
\end{equation}
The quantum supercharges are obtained from the classical ones by Weyl ordering procedure \cite{howto}. This gives
(\ref{eq_Qf}). The anticommutator $\{Q_\alpha, \bar Q^\alpha\}$ gives the quantum Hamiltonian  (\ref{eq_susyham}). 

As was noticed, the field $A_\mu$ naturally obtained in the HSS framework satisfies the constraint  $\partial_\mu \A_\mu=0$
\cite{Ivanov:2003nk}.
This does not really impose a restriction, however, because gauge transformations of $A_\mu$ that shift it by 
the gradient of an arbitrary
 function amount to adding a total derivative in the Lagrangian (\ref{43}).

\section*{Acknowledgments}

We are indebted to Evgeny Ivanov for illuminating discussions.
The work of MK was supported in part by grants RFBR-07-02-01161 and
grant of leading scientific schools NSH-3036.2008.2.

\appendix

\section*{Appendix: Harmonic superspace in quantum mechanics}\label{sec_susp}

\renewcommand{\theequation}{A\arabic{equation}}
\setcounter{equation}{0}

In this appendix, we introduce some basic HSS notations and definitions (see Ref.~\cite{Galperin:2001uw} for detailed explanations)
in application to quantum mechanical systems.

Consider the  ordinary  $\ca N=2$ superspace 
\begin{equation}
  \mathbb{R}^{1|4}=\left\{t,\theta_\alpha,\bar\theta^\beta\right\} \ ,
\end{equation}
with 
$\theta_\alpha$ and $\bar \theta^\beta = \varepsilon^{\beta\gamma} \bar \theta_\gamma = (\theta_\beta)^\dagger$ belonging 
 to the fundamental representation of ${\rm SU}(2)$.

Introduce the supercharges
\footnote{Our convention follows the convention in Ref.~\cite{IvSmil}, but differs from 
the convention of Ref.~\cite{Ivanov:2003nk} by the change of time direction $t\rightarrow -t$.
With this, we reproduce the  correct sign in the kinetic term for the spinor field in Eq.~(\ref{eq_41}).}
\begin{equation}
  Q^\alpha=\pop{\theta_\alpha}+i\bar\theta^\alpha\pop{t} , \vergule
  \bar Q_\alpha=\pop{\bar\theta^\alpha}+i\theta_\alpha \pop{t}
\end{equation}
and superderivatives
\begin{equation}\label{eq_superder}
  D^\alpha=\pop{\theta_\alpha}-i\bar\theta^\alpha\pop{t} , \vergule
  \bar D_\alpha=\pop{\bar\theta^\alpha}-i\theta_\alpha \pop{t}.
\end{equation}
The supercharges form the   $\ca N=2$  SUSY algebra,  while the superderivatives anticommute with $Q^\alpha$ and $\bar Q_\beta$, 
\begin{equation}
  \left\{Q^\alpha,\bar Q_\beta\right\}=2\delta^\alpha_\beta\, i\partial_t,\vergule
    \left\{D^\alpha,\bar D_\beta\right\}=-2\delta^\alpha_\beta\, i\partial_t.
\end{equation}

  To proceed to harmonic superspace $\mathbb{H}\mathbb{R}^{1+2|4}=\mathbb{R}^{1|4}\times {\rm S}^2$,  we introduce a set of 
two complex coordinates $u^{+\alpha}$. Introduce also $u^-_\alpha = (u^{+\alpha})^*$ and impose the condition
  \begin{equation}
\label{norma}
  u^{+\alpha} u^-_\alpha=1 \ .
\end{equation}
Then $u^{+\alpha}$ parametrize the R-symmetry group  ${\rm SU}(2)$. 
The differential operators 
 \begin{equation}\label{eq_Dpp}
  D^{++}=u_\alpha^+\pop{u_\alpha^-}, \ \ \ \ \ D^{--}   =  u_\alpha^-\pop{u_\alpha^+}\ , \ \ \ \ \ \ 
D^0 = u_\alpha^+\pop{u_\alpha^+} - u_\alpha^-\pop{u_\alpha^-}
\end{equation}
 are called {\it harmonic derivatives}. The ${\rm U}(1)$ charge operator $D^0$ plays a special role. The functions of zero ${\rm U}(1)$ charge
live on the coset  ${\rm S}^2 =  {\rm SU}(2)/{\rm U}(1)$. The coordinates $u^+_\alpha$ have charge 1,
 the coordinates $u^-_\alpha$ have charge -1,
etc. 

One can define now harmonic projections  $D^\pm=u^\pm_\alpha D^\alpha$, $\bar D^\pm=u^\pm_\alpha \bar D^\alpha$.
It is convenient to go over in the  {\sl analytic basis} in HSS,
\begin{equation}
  \mathbb{H}\mathbb{R}^{1+2|4}=\left\{t_{\rm A}, \theta^\pm,\bar\theta^\pm,u^\pm_\alpha\right\},
\end{equation}
where
\begin{equation}\label{eq_analcoord}
  t_{\rm A}=t+i\left(\theta^+\bar\theta^-+\theta^-\bar\theta^+\right),\vergule
  \theta^\pm=u^\pm_\alpha\theta^\alpha ,\vergule
  \bar\theta^\pm=u^\pm_\alpha\bar\theta^\alpha.
\end{equation}
In this basis, the covariant spinor derivatives $D^+,\ \bar D^+$ are just 
\begin{equation}\label{eq_Dana}
  D^+=\pop{\theta^-},\vergule \bar D^+=-\pop{\bar\theta^-}\ ,
\end{equation}
while the operator $D^{++}$ acquires the  form
\begin{equation}\label{eq_Dppana}
  D^{++}=u^+_\alpha\pop{u^-_\alpha}+\theta^+\pop{\theta^-}+\bar\theta^+\pop{\bar\theta^-}
    +2i\theta^+\bar\theta^+ \pop{t_{\rm A}} \ .
\end{equation}

The derivative operators $D^+$, $\bar D^+$, $D^{++}$ (anti)commute with each other and with supercharges. 
Because of this, it is possible to consider a superfield $q^+$ with ${\rm U}(1)$ charge +1 satisfying
\begin{equation}\label{eq_anal}
  D^+ q^+=0,\vergule
  \bar D^+ q^+=0,\vergule
  D^{++}q^+=0.
\end{equation}
In the analytic superspace coordinates, the first and the second equations mean
 that $q^+$ depend only on $\theta^+$ and $\bar\theta^+$, but not on $\theta^-$ and $\bar\theta^-$. 
This is the so-called {\sl superfield analyticity condition}. When expanding 
the field $q^+(t_A, \theta^+, \bar\theta^+, u^\pm_\alpha)$ over spinor coordinates and the harmonics, one obtains an infinite set
of physical fields $\Phi(t_A)$. However,  imposing also the condition $D^{++} q^+ = 0$ 
drastically reduces the number of such fields, making it finite. In the analytic basis, 
the solution of the constraints (\ref{eq_anal}) reads 
   \begin{equation}\label{eq_q}
  q^+=x^\alpha(t_{\rm A})u^+_\alpha
  -2\theta^+\chi(t_{\rm A})-2\bar\theta^+\bar\chi'(t_{\rm A})
  -2i\theta^+\bar\theta^+\partial_{\rm A}x^\alpha(t_{\rm A}) u^-_\alpha
\end{equation}
with the factors $-2$ introduced for convenience.

The constraints (\ref{eq_anal}) admit an  involution symmetry $q^+\rightarrow \widetilde {q^+}$ which commutes with SUSY transformations 
\cite{Ivanov:2003nk,Galperin:2001uw}.
This involution acts just as the ordinary complex conjugation {\it except} its action on the harmonics $u^\pm_\alpha$, which is 
\begin{equation}
\label{harminv}
\widetilde {u^\pm_\alpha}=u^{\pm \alpha},\vergule
  \widetilde {u^{\pm \alpha}}=-u^\pm_\alpha \ .
\end{equation}
This gives
\begin{equation}
\label{eq_involution}
  \widetilde{t_{\rm A}}=t_{\rm A},\vergule
  \widetilde {\theta^\pm}=\bar\theta^\pm,\vergule
  \widetilde {\bar\theta^\pm}=-\theta^\pm,
\end{equation}
and hence 
 \begin{equation}\label{eq_tildeq}
 \widetilde{q^+}=\left[x_\alpha(t_{\rm A})\right]^*u^+_\alpha
    -2\theta^+\bar\chi'^*(t_{\rm A})+2\bar\theta^+\chi^*(t_{\rm A})
    -2i \theta^+\bar\theta^+\partial_{\rm A}\left[x_\alpha(t_{\rm A})\right]^* u^-_\alpha .
\end{equation}
It is straightforward to see  that the field  $\widetilde {q^+}$ satisfies the same constraints (\ref{eq_anal}) as the field $q^+$.
The involution symmetry was used in the main text to impose the pseudoreality condition (\ref{reality}) 
on the field $q^{+\dot\alpha}$.

The invariant actions involve the harmonic integral $\int du$. To find such integral of any function $f(u^\pm_\alpha)$, one should  
expand $f$ in the harmonic Taylor series and, for each term, do the integrals using the rules 
\begin{equation}
\label{intharm}
\int du\, 1=1,\vergule \int du\, u^+_{\{\alpha_1}\dots u^+_{\alpha_k}u^-_{\alpha_{k+1}}\dots u^-_{\alpha_{k+\ell}\}}=0 \ ,
\end{equation}
where the integrand is  symmetrized over all indices.
The values of the integrals of all other harmonic monoms (for example, $\int du \, u^+_\alpha u^-_\beta = \frac 12 \varepsilon_{\alpha\beta}$) follow from
(\ref{intharm}) and the definition (\ref{norma}).

\end{document}